\documentclass{article}
\usepackage{spconf,amsmath,graphicx}
\usepackage{multirow}
\usepackage{comment}
\usepackage{cite}
\usepackage{array}
\usepackage{amsmath,amssymb,amsfonts}
\usepackage{graphicx}
\usepackage{textcomp}
\usepackage{subfigure} 
\usepackage{xcolor}
\usepackage{color}
\usepackage{hyperref}
\usepackage{multirow}
\usepackage[algoruled,vlined,linesnumbered]{algorithm2e}
\usepackage{bm}
\usepackage{booktabs}
\usepackage{dsfont}
\usepackage{bbding}
\usepackage{newfloat}
\usepackage{listings}
\usepackage{multirow}
\usepackage{makecell}
\usepackage[colorinlistoftodos]{todonotes}
\usepackage{tikz,xcolor,hyperref}

\title{A Meta-Learning Based Gradient Descent Algorithm for \\MU-MIMO Beamforming}

\name{Jing-Yuan Xia$^{1}$, Zhixiong~Yang$^{1}$, Tong~Qiu$^{1}$, Huaizhang Liao$^{1}$ and Deniz~Gündüz$^{2}$
}
\address{$^{1}$College of Electronic Engineering, National University of Defense Technology, China\\
$^{2}$Department of Electrical and Electronic Engineering, Imperial College London, UK}

\begin{document}

\maketitle

\begin{abstract}

Multi-user multiple-input multiple-output (MU-MIMO) beamforming design is typically formulated as a non-convex weighted sum rate (WSR) maximization problem that is known to be NP-hard. This problem is solved either by iterative algorithms, which suffer from slow convergence, or more recently by using deep learning tools, which require time-consuming pre-training process. In this paper, we propose a low-complexity meta-learning based gradient descent algorithm. A meta network with lightweight architecture is applied to learn an adaptive gradient descent update rule to directly optimize the beamformer. This lightweight network is trained during the iterative optimization process, which we refer to as \emph{training while solving}, which removes both the training process and the data-dependency of existing deep learning based solutions. 
Extensive simulations show that the proposed method achieves superior WSR performance compared to existing learning-based approaches as well as the conventional WMMSE algorithm, while enjoying much lower computational load. 
\end{abstract}


\maketitle

\section{Introduction}\label{sec:introduction}

Downlink beamforming design is essential for the communication over multi-antenna cellular networks. In multi-user multiple-input multiple-output (MU-MIMO) systems \cite{xu2017waveforming}, the transmit beamforming design that maximizes the weighted sum rate (WSR) with respect to a total power constraint entails solving an NP-hard non-convex problem \cite{luo2008dynamic}. The weighted minimum mean square error (WMMSE) algorithm introduced in \cite{shi2011iteratively} is the most widely used approach to obtain a suboptimal solution. While it does provide satisfactory WSR performance via an alternating minimization (AM) approach \cite{christensen2008weighted, schmidt2009minimum, shi2011iteratively, luo2019joint, zhang2020robust}, it has high computational complexity due to matrix inversion and bisection search, which limits its application in practical scenarios.

Recently deep learning based black-box approaches have gained popularity to address the complexity of the iterative solutions for MU-MIMO beamforming \cite{sun2018learning, xia2019deep, kim2020deep, zhang2021model, chowdhury2021unfolding}. The complexity of these solutions depend on the network architecture, and bulk of the complexity is transferred to the training process. However, these methods have the following limitations: i) when trained in a supervised manner, the performance is bounded by the training data, which is typically generated by the WMMSE algorithm; ii) poor interpretability on algorithmic principles, which limits the feasibility of incorporating expert knowledge; and iii) lack of generalization, which means time-consuming retraining or tuning is always necessary when the communication environment changes. More recently, \emph{deep unfolding} based approaches \cite{pellaco2021deep, chowdhury2021unfolding, gao2018enhanced, li2020deep} have been proposed for model explainability and reduction in computational complexity. Deep unfolding provides a layer-wise explainable deep network architecture with respect to the reference iterative algorithm, e.g., the WMMSE algorithm; and attains a trade-off between computational complexity and performance. The latest works \cite{hu2020iterative} and \cite{pellaco2021deep} unfold the WMMSE algorithm for the MIMO beamforming problem with significantly less computational complexity; however, their performance is upper bounded by the WMMSE algorithm, and re-training is typically necessary and time-consuming for different environments, such as variations in the number of users, or the number of transmit and receive antennas. To address these issues, in \cite{xia2022metalearning, xia2021meta, yang2022learning}, we proposed a meta-learning based beamforming design approach for the multiple input- single output (MISO) setting, which leads to an adaptive gradient descent (GD) strategy for beamformer estimation, and showed that it surpasses the performance of the WMMSE algorithm, particularly in the high signal-to-noise ratio (SNR) regime.


In this paper, we propose a meta-learning based gradient descent (MLGD) algorithm for solving arbitrary MU-MIMO tasks. Applying the meta-learning approach to the MU-MIMO problem is not a straightforward extension as the update of the weight matrices imposes positive semi-definiteness constraints, and the multi-user problem has a more challenging optimization geometry. The main contributions of this paper are summarized next: i) Instead of converting the WSR maximization problem into the WMMSE reformulation that entails AM corresponding to the beamformer, weight matrices, and the receiver gain, the proposed MLGD approach strives to maximize the WSR directly by optimizing the beamformer. In this way, the matrix inversion and the positive semi-definiteness constraints are avoided, thus allowing a lightweight network to learn flexible gradients that can adaptively optimize the beamformer matrices in an iterative fashion. ii) Thanks to the limited computational requirements of training the lightweight network, the proposed MLGD algorithm essentially follows the \emph{training while solving} approach, which updates its network parameters during the iterative optimization process of the beamformer. Therefore, the MLGD algorithm offers high efficiency and flexibility for MU-MIMO beamforming in a variety of scenarios as it does not require any pre-training or fine-tuning. 
iii) We show through simulation results that the MLGD algorithm surpasses the WMMSE algorithm as well as  the existing learning based approaches in terms of performance achieved by the WSR; and it is faster and requires less computations than the existing learning-based approaches thanks to its lightweight network architecture. Finally, it allows flexible implementation on arbitrary MU-MIMO scenarios in both the fully and lightly loaded settings. 


\section{Problem formulation}

\subsection{System Model}
We consider a MU-MIMO downlink channel, where the transmitter with $N_t$ antennas serves $K$ users, each equipped with $N_r$ antennas and sends $d$ independent data streams to $K$ users. The beamformer $\bm{V}_k\in\mathbb{C}^{N_t\times d}$ is used to precode the information sent to user $k$. Then the signal received at the $k^{th}$ user, $k \in \mathcal{K} \triangleq \{1, \ldots, K\}$, is given by
\begin{equation}
\bm{y}_k = \bm{H}_{k} \bm{V}_k \bm{s}_k+\sum_{r\in \mathcal{K}, r\neq k}\bm{H}_{k} \bm{V}_r \bm{s}_r + \bm{n}_k, \label{eq-signal received}
\end{equation}
where $\bm{H}_{k}\in \mathbb{C}^{N_t \times N_r}$ is the channel to the $k^{th}$ user, $\bm{s}_k\in\mathbb{C}^{d}$ is the signal vector destined for the $k^{th}$ user, and $\bm{n}_k\in\mathbb{C}^{N}$ is the additive white Gaussian noise at user $k$. Let $\mathcal{V}=(\bm{V}_{1}, \ldots, \bm{V}_{K})$ denote the sequence of all the beamformers.


Let ${R}_k$ denote the rate of user $k$ defined as follows, 
 	\begin{equation}
        R_k \triangleq \log \det{(\bm{I} + \mathcal{M}_k 
        (\sum_{r\neq k} \mathcal{M}_r + \sigma_k^2\bm{I})^{-1})}.
 		\label{eq-R_k}
     \end{equation}
where$\mathcal{M}_j=\bm{H}_{k} \bm{V}_j \bm{V}_j^H \bm{H}_{k}^H$ and $P$ is the maximum total transmit power. 
The primary WSR maximization problem can be written as
\begin{equation}\label{eq-maxWSR}
\begin{aligned}
\max_{\mathcal{V}} & F(\mathcal{V})=\sum_{k=1}^{K} \alpha_{k} R_{k} \\
\text { s.t. } & \sum_{k \in \mathcal{K}} \operatorname{Tr}\left(\boldsymbol{V}_{k} \boldsymbol{V}_{k}^{H}\right) \leq P,
\end{aligned}
\end{equation}
where $\alpha_k$ is the weight of user $k$.  

 


\subsection{Existing Solutions}
A generic GD based solution can be employed to solve the WSR problem in (\ref{eq-maxWSR}), in which the variables $\bm{V}_k$, $k \in \mathcal{K}$, are optimized in an iterative fashion. The common GD based update rule at the $i^{th}$ iteration can be written as
\begin{equation}
\bm{V}_{k,i+1} = \bm{V}_{k,i} + \gamma_{i}\cdot \text{g}(\nabla_{\bm{V}_{k,i}}F(\mathcal{V})), \label{eq-VGD}
\end{equation}
where $\nabla_{\bm{V}_k}F(\left\{ \bm{V}_k \right\}_{ k \in \mathcal{K}})$ is the gradient of the WSR with respect to the current beamformer, $\text{g}(\cdot)$ denotes a hand-crafted variable update function, such as momentum, RMSprop or Adam \cite{kingma2014adam}, and $\gamma_{i}$ represents the step size at the $i^{th}$ iteration. 
However, due to the dramatic non-convexity of the problem (\ref{eq-maxWSR}), the first-order gradient descent solutions typically converge to bad local optima, where the gradient vanishes, i.e., $\nabla_{\bm{V}_k}F(\mathcal{V})=0$.

As such, the WMMSE algorithm was proposed in \cite{shi2011iteratively} that reformulates the non-convex and NP-hard problem (\ref{eq-maxWSR}) in the following form
\begin{equation} \label{WMMSE-WSR}
\begin{aligned}
\min _{\left\{\boldsymbol{U}_{k}, \boldsymbol{W}_{k}, \boldsymbol{V}_{k}\right\}_{k \in \mathcal{K}}} &\sum_{k=1}^{K} \alpha_{k}\left\{\operatorname{Tr}\left(\boldsymbol{W}_{k} \widetilde{\boldsymbol{E}}_{k}\right)-\log \operatorname{det}\left(\boldsymbol{W}_{k}\right)\right\}\\
\text{s.t.}  & \sum_{k=1}^{K} \operatorname{Tr}\left(\boldsymbol{V}_{k} \boldsymbol{V}_{k}^{H}\right) \leq P , 
\end{aligned}
\end{equation}
where $\tilde{\bm{E}_k} = \tilde{\bm{E}_k} \left( \bm{U}_k,\left\{ \bm{V}_k \right\}_{ k \in \mathcal{K}} \right)$ is the MSE matrix of user $k$, $\bm{U}_k$  and $\bm{W}_k\in\mathbb{C}^{d\times d}$ is a positive semi-definite weight matrix for user $k$. This problem is convex in each individual variable, and the WMMSE algorithm carries an alternating minimization (AM) strategy that sequentially optimizes each variable with respect to minimizing an amenable sub-problem.

However, the WMMSE algorithm still suffers from the high computational complexity due to the bisection search and matrix inversion, as well as the performance limitation caused by the intrinsic non-convexity (i.e., it is only guaranteed to converge to a stationary point, not necessarily the global optima). To reduce the computational complexity, deep learning and deep unfolding techniques have been introduced to replace the matrix inversion procedures and parameter tuning by network-based trainable modules. 
Despite achieving significant reduction in the computational complexity, achieved WSR by these learning-based methods typically falls short of the one achieved by the WMMSE algorithm, while they also require significant pre-training time. While this additional training cost can be ignored if the same parameters can be used forever, this can hardly be the case in a mobile wireless network where the number of users, their antenna numbers and the environment statistics can change quite often. This creates a significant limitation on the applicability of pre-trained neural network solutions in practical scenarios.

\begin{algorithm}[t]
    \SetAlgoLined
    \textbf{Given:} Number of antennas $N_t$ and $N_r$, number of users $K$,  update index set $\mathcal{S}$.
    
    \textbf{Initialize:} $\bm{\theta}_{0}$, $\mathcal{V}_0$, ${\left\{\boldsymbol{H}_{k}\right\}_{k \in \mathcal{K}}}$, $\mathcal{L}_{meta}=0$.\\
    \For{i$\gets$ 1, 2,$\ldots$, I}{
        \For{k$\gets$ 1, 2,$\ldots$, K}{

        
            $\Delta \bm{V}_{k,i} = \text{G}_{\bm{\theta}_{s}}(\nabla_{\bm{V}_{k,i}}F(\mathcal{V}_i))$
        
            $\bm{V}_{k,i+1} = \bm{V}_{k,i} + \Delta \bm{V}_{k,i}$
            
            $\bm{V}_{k,i+1} = \Omega(\bm{V}_{k,i+1})$
        }
    $\mathcal{L}_{meta}=\mathcal{L}_{meta}+F(\mathcal{V}_i)$
        
        \If{$i\in \mathcal{S}$}{   
                  $\bm{\theta}_{s+1} =\bm{\theta}_{s}+ \gamma \cdot\mathrm{Adam}(\nabla_{\bm{\theta}_{s}}\mathcal{L}_{meta})$ 
                  
                  $\mathcal{L}_{meta}=0$
          }

    }
    \textbf{Output:} $\mathcal{V}_I$.
\caption{\label{alg1}MLGD algorithm for WSR maximization
}
\end{algorithm}

\section{Proposed Solution}
We propose a learning-aided but training-free MLGD algorithm to directly optimize the beamformer for a given network setting. The MLGD algorithm is applied to the non-convex problem (\ref{eq-maxWSR}) instead of its variant in (\ref{WMMSE-WSR}), as such the cumbersome AM with matrix inversion step is bypassed by an iterative learning-based gradient descent loop.  Let $\mathcal{V}_{i}=[\bm{V}_{1,i}, \bm{V}_{2,i},\cdots, \bm{V}_{K,i}]$ denote the beamformers of all $K$ users at the $i^{th}$ iteration.
The MLGD algorithm updates the beamformers $\bm{V}_{k,i}$ via a function parametrized by a lightweight neural network, whose parameters are updated in the way of meta-learning strategy to learn a flexible update rule with a bird eye on the past optimization trajectory. 

Mathematically, a lightweight network, denoted by $\text{G}_{\bm{\theta}_{s}}(\cdot)$, with parameters $\bm{\theta}_{s}$, dynamically and adaptively learns to optimize the transmit beamformer $\bm{V}_{k, i}$. The network $\text{G}_{\bm{\theta}_{s}}(\cdot)$  takes as input the gradient $\nabla_{\bm{V}_{k,i}}F(\mathcal{V}_{i})$, and outputs the term to update the beamformer in the following form
\begin{equation}\label{eq-LAGD-Net} 
    \bm{V}_{k,i+1} = \bm{V}_{k,i} +\text{G}_{\bm{\theta}_{s}}(\nabla_{\bm{V}_{k,i}}F(\mathcal{V}_{i})).
\end{equation}
We note that the input of the network contains the beamformers of all the users, so that the $k^{th}$ beamformer $\bm{V}_{k,i}$ is updated with respect to the gradient of the WSR of all the users $\nabla_{\bm{V}_{k,i}}F(\mathcal{V}_{i})$. This essentially incorporates the inter-user relations for the individual users' beamformer optimization.   
In Equation (\ref{eq-LAGD-Net}), the network parameters $\bm{\theta}_{s}$ determine the update rule, which is continuously optimized by back-propagating the WSR value in the objective function in Equation (\ref{eq-maxWSR}).\footnote{We emphasize that the network $\text{G}_{\bm{\theta}}$ starts from scratch, i.e., random initialization, and is trained during the iterations of the solution algorithm (\ref{eq-maxWSR}).} The \textit{training while solving} approach of the MLGD algorithm provides a data-driven approach that can also provide computational simplicity and practical flexibility. 

The MLGD algorithm is applied in a plug-and-play fashion, that has no prior training requirement. The parameters of the neural network, $\bm{\theta}_{s}$, are updated by the Adam \cite{kingma2014adam} optimizer in a meta-learning strategy with respect to the leveraged meta loss following the concept in \cite{xia2022metalearning}:
\begin{equation} \label{eq-MLGD-optimizer}
    \bm{\theta}_{s+1}=\bm{\theta}_{s}+\gamma\cdot\mathrm{Adam}(\nabla_{\bm{\theta}_{s}}\mathcal{L}_{meta}),
\end{equation}
where $\mathcal{L}_{meta}=\sum_{i=(s-1)T+1}^{sT}F(\mathcal{V}_{i})$ is the leveraged meta loss and $\gamma$ denotes the learning rate. In (\ref{eq-MLGD-optimizer}), $s$ corresponds to the network update index, with $s=1,2,\ldots,I/T$, where $I$ is the total number of iterations and $T$ is the update interval referring to the number of iterations to be leveraged for meta-learning. We define the update index set $\mathcal{S}=\left\{T,2T,\ldots,K\right\}$ to indicate that the meta-learning strategy updates the network parameters at every $T$ iterations. In this way, the learned optimization rule has a bird eye view of the global behaviour of the optimization trajectory taking into account the interactions between the rates of all the users, instead of exhaustively optimizing the beamformer based on the local view of the WSR value $F(\mathcal{V}_{i})$. This intrinsically ensures that the MLGD is capable of circumventing bad local optimum, thus achieving better WSR performance.  
To satisfy the total power constraint in (\ref{eq-maxWSR}), the beamformers obtained from (\ref{eq-LAGD-Net}) are projected at each step by $\Omega(\bm{V}_k) =  \sqrt{\frac{P}{\sum_{k=1}^{K} \operatorname{Tr}\left(\boldsymbol{V}_{k} \boldsymbol{V}_{k}^{H}\right)}} \bm{V}_k, \text{for } k \in \mathcal{K}$.

We highlight that MLGD is capable of escaping from bad local optima. This can be realized by noticing that even if the $i^{th}$ gradient $\nabla_{\bm{V}_{k,i}}F(\mathcal{V}_{i})=0$, it does not invalidate the update for $\bm{V}_{k,i}$ in light of the trained parameters $\bm{\theta}_s$, and the corresponding WSR will be accumulated to update $\bm{\theta}_s$ to achieve a higher WSR value. This ensures the direct optimization of the problem (\ref{eq-maxWSR}) rather than iteratively optimizing local problems, which can miss the direction of the global objective function. The general structure of the proposed MLGD algorithm is presented in \textbf{Algorithm} \ref{alg1}. 

Based on the outlined algorithmic description, the key advances of the MLGD algorithm with respect to the existing solutions are: i) MLGD optimizes $\bm{V}$ directly via solving problem (\ref{eq-maxWSR}) with GD, which enjoys less computational complexity, compared to the alternatives (e.g., WMMSE) that involve computationally demanding matrix-inversion operations and bisection search; ii) The incorporated meta-learning approach directly updates the beamformers with the global objective in mind, which allows MLGD to avoid local optima and have the capability to outperform existing solutions, including WMMSE; and iii) MLGD is training-free thanks to the few-shot learning ability of the devised meta-learning approach, and the lightweight architecture of $G_{\theta}$ allows \textit{training while solving}. Existing deep learning and deep-unfolding methods rely on a time-consuming and data-hungry pre-training process every time the environment statistics change.


\section{Simulation Results}
All the simulated algorithms are implemented in Python 3.7 with Pytorch 1.8. 
The full code is available at \cite{MLGD2022code}.
All the presented results are averaged of $10^5$ realizations of the channel ${\left\{\boldsymbol{H}_{k}\right\}_{k \in \mathcal{K}}}$ with Rayleigh distributed channel based on the complex Gaussian distribution $\mathcal{CN}(0,1)$.
The WMMSE algorithm is applied as the baseline with the stopping criteria that the variation in WSR between two consecutive iterations is less than or equal to $10^{-4}$ bits. 
The recent deep-learning based state-of-the-art method, algorithm from \cite{hu2020iterative}, is also considered for comparison, that employs an end-to-end deep network as optimizer. 
We set $\alpha_k=1$ for all the users. The learning rate of the Adam optimizer for network update is set to $5\times10^{-4}$. A lightly load case with ($N_t=8$, $N_r=2$, $d=2$, $K=2$), and a fully load case with ($N_t=8$, $N_r=2$, $d=2$, $K=4$) are considered.
A lightweight fully-connected network (FCN) is applied to the network $\text{G}_{\theta}(\cdot)$, which employs 2 hidden layers, with 50 nodes at each.
We randomly initialize the algorithms for 10 times and pick the best results for all approaches.
\begin{figure}[t]
	\centering
	\hspace*{\fill}
	\subfigure[\textcolor{black}{WSR vs. SNR}]{
 		\label{fig-SNR2}
		\includegraphics[width=0.46\linewidth]{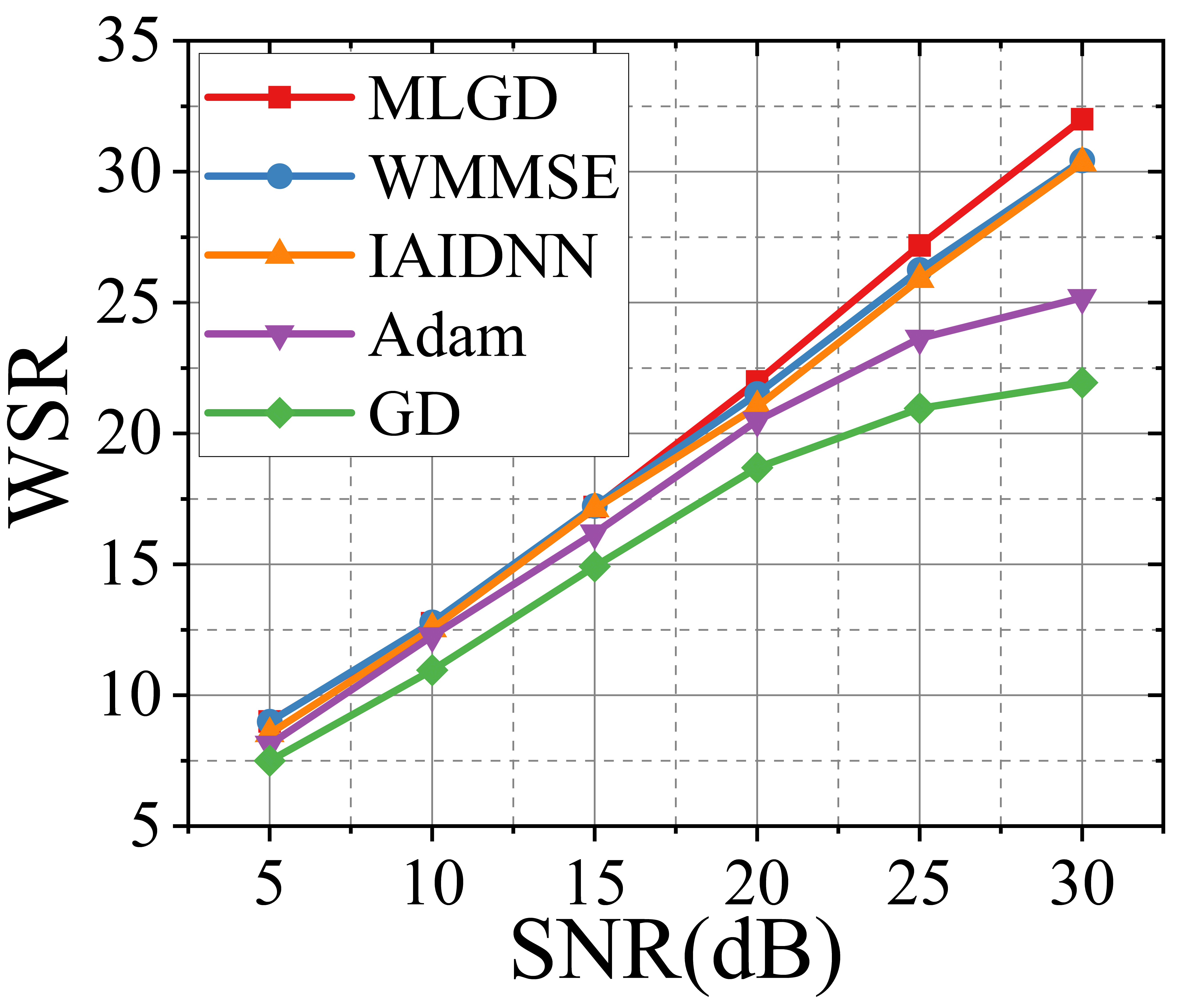}}
		\hfill
	\subfigure[Variance of WSR values]{
        \label{fig-variance2}
		\includegraphics[width=0.46\linewidth]{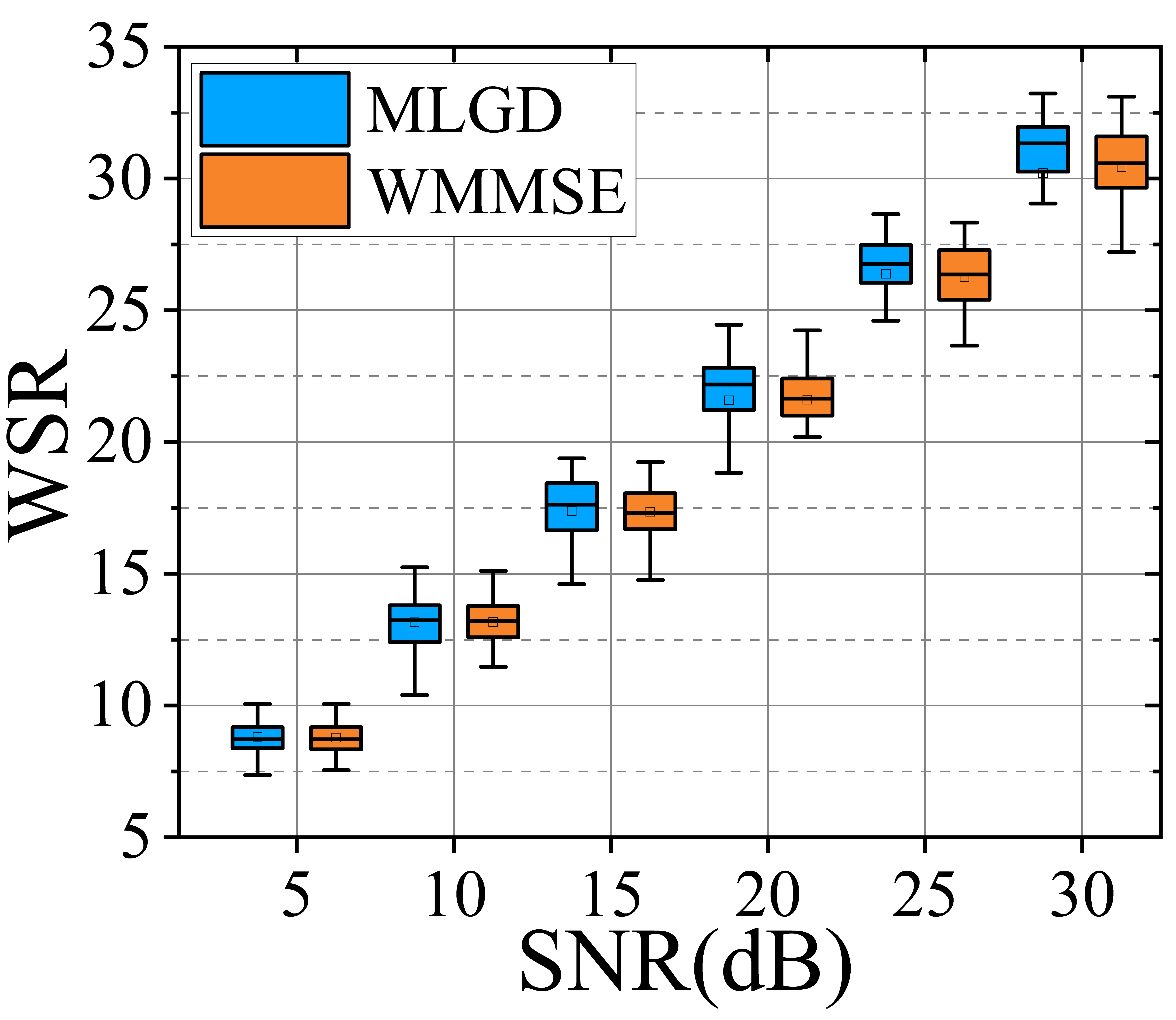}}
	\hspace*{\fill}
	\caption{\textcolor{black}{Compared with WMMSE \cite{shi2011iteratively}, IAIDNN \cite{hu2020iterative} and gradient-based conventional GD and Adam \cite{kingma2014adam} schemes. ($N_t=8, K=N_r=d=2$).}}
\end{figure}

\begin{figure}[t]
	\centering
	\hspace*{\fill}
	\subfigure[\textcolor{black}{WSR vs. SNR}]{
 		\label{fig-SNR4}
		\includegraphics[width=0.46\linewidth]{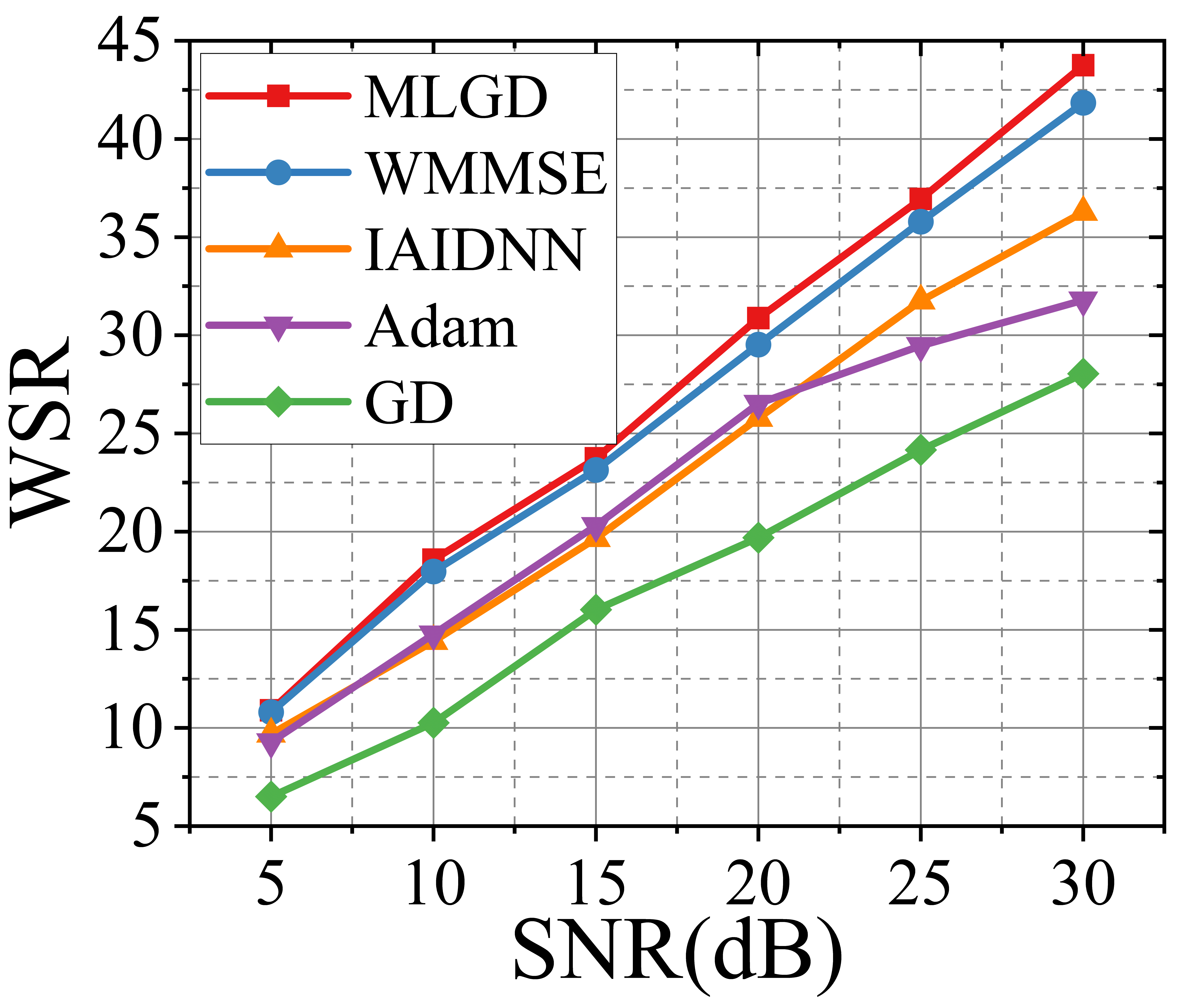}}
		\hfill
	\subfigure[Variance of WSR values]{
        \label{fig-variance4}
		\includegraphics[width=0.46\linewidth]{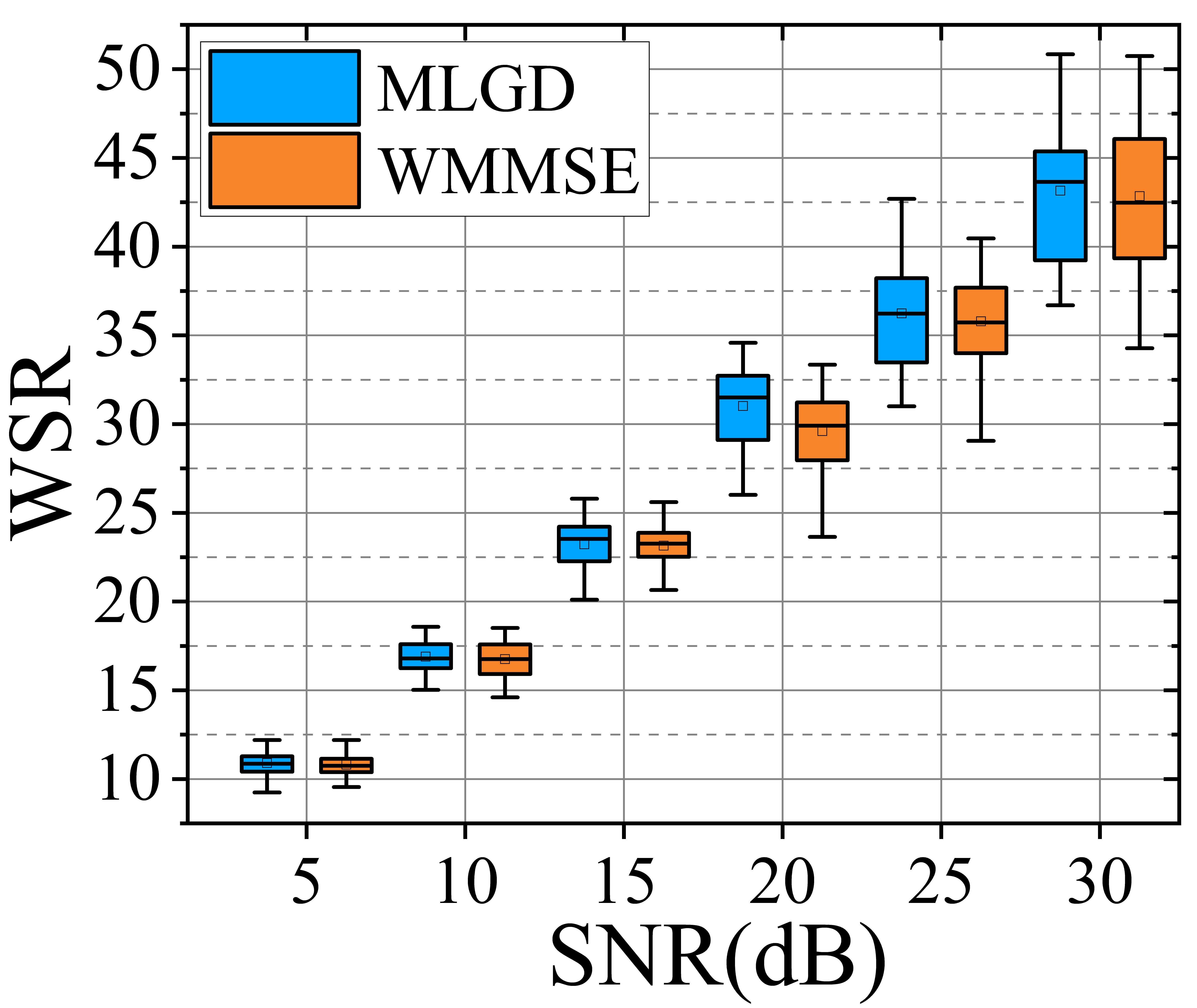}}
	\hspace*{\fill}
	\caption{\textcolor{black}{Compared with WMMSE \cite{shi2011iteratively}, IAIDNN \cite{hu2020iterative} and gradient-based conventional GD and Adam \cite{kingma2014adam} schemes. ($N_t=8, N_r=d=2, K=4$).}}
\end{figure}

In Fig. \ref{fig-SNR2}, the curves of the WSR results of the proposed MLGD with the widely-used WMMSE \cite{shi2011iteratively}, IAIDNN \cite{hu2020iterative}, standard GD and Adam \cite{kingma2014adam} approaches for a lightly loaded scenario are presented. The proposed MLGD algorithm significantly surpasses the WMMSE and IAIDNN in high SNR regime of SNR=25-30dB, and achieves competitive performance when SNR=5-20dB. This can be understood that the non-convexity of the WSR problem grows with the SNR, and the meta-learning based optimization strategy can attain better optima than the WMMSE algorithm.  
In Fig. \ref{fig-variance2}  this claim is further verified by the results from a fully loaded case with $K=4$, where the gaps between the MLGD and the comparing approaches become wider. We argue that with the raise of the network load the non-convexity of the WSR problem increases, therefore leading to more challenging optimization in AM fashion as the WMMSE and its variants can be stuck at saddle points or bad local optima more frequently.  
In Fig. \ref{fig-SNR4} and Fig. \ref{fig-variance4}, the variances of the results from the proposed MLGD and the baseline WMMSE algorithm depicts that MLGD enjoys better stability than the WMMSE algorithm. Especially in high SNR regime, it is apparent that the MLGD has a smaller variance with better upper and lower value than the WMMSE algorithm. This is also validates that MLGD can avoid bad local optima that the WMMSE may get stuck when starting from one certain channel realization.


\begin{table}[thpb]  \caption{Comparison of computational requirements} 
 \label{computational complexity}
\renewcommand{\arraystretch}{1}
\setlength{\tabcolsep}{1pt}
\begin{center}
 \begin{tabular}{llll}  
\toprule   
  Methods                        &Train     &\;\;Test Complexity             &Model Size       \\  
\midrule   
  \textbf{MLGD algorithm }                     & \XSolidBrush       & $\mathcal{O}(TKN_t^2N_r + TN_r^3)$            & $\sim 10^{2}$                 \\
  IAIDNN \cite{hu2020iterative}                & \checkmark         & $\mathcal{O}(KN_t^2N_r)$       & $\sim 10^{3}$                 \\  
  Adam \cite{kingma2014adam}/ GD               & \XSolidBrush       & $\mathcal{O}(TKN_t^2N_r + TN_r^3)$            & \XSolidBrush             \\
  WMMSE \cite{shi2011iteratively}              & \XSolidBrush       & $\mathcal{O}(TKN_t^2N_r + TN_r^3)$           & \XSolidBrush             \\
  \bottomrule  
\end{tabular}
\end{center}
\end{table}

In Table \ref{computational complexity}, the  where $T$ denotes the required number of iterations. Looking at the test computational complexity in terms of $N_r$ and $N_t$, MLGD has a similar complexity comparing to the WMMSE. However, the majority of matrix inversion operations in the WMMSE are substituted by network-based forward behavior in MLGD, and thus the complexity reduces dramatically in practical. Meanwhile, though the IAIDNN enjoys lower test complexity, the necessary training in advance significantly arises the burden on practical implementation in varying scenarios. 


\section{Conclusions}
In this paper, we proposed a meta-learning based gradient descent (MLGD) algorithm for MU-MIMO beamforming. In particular, a meta-learning scheme is proposed for optimizing the transmit precoder in a less-greedy and more effective way. Different from the existing learning-based approaches, MLGD can be applied in a plug-and-play fashion as no pre-training is required.  Convergence and monotonicity of the MLGD will be analyzed in future works. 




\newpage
\newpage

\bibliographystyle{IEEEtran}
\addcontentsline{toc}{section}{\refname}\bibliography{Bib_MLBF}

\end{document}